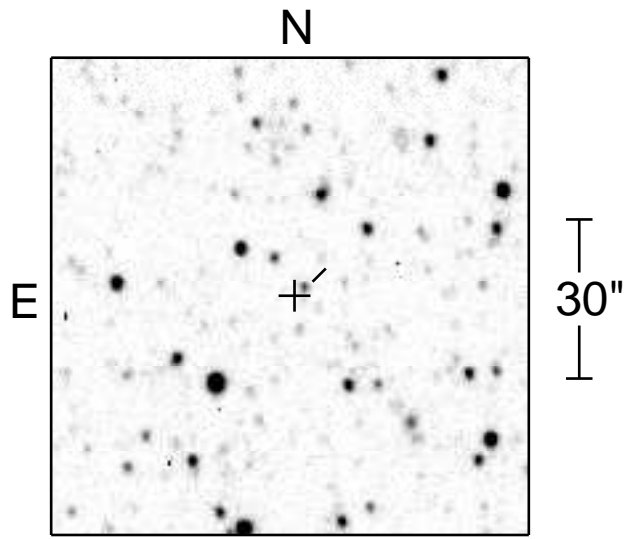

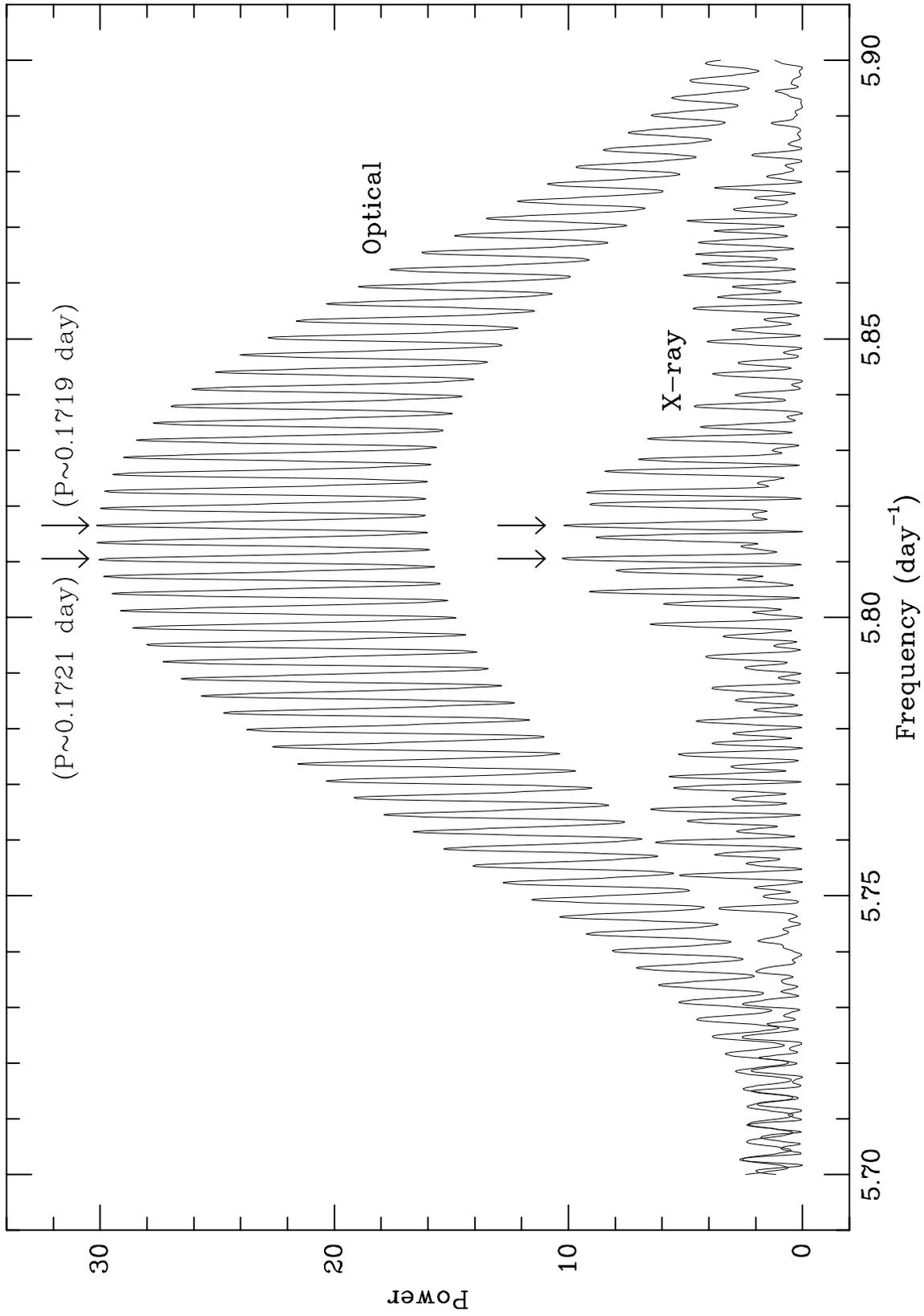

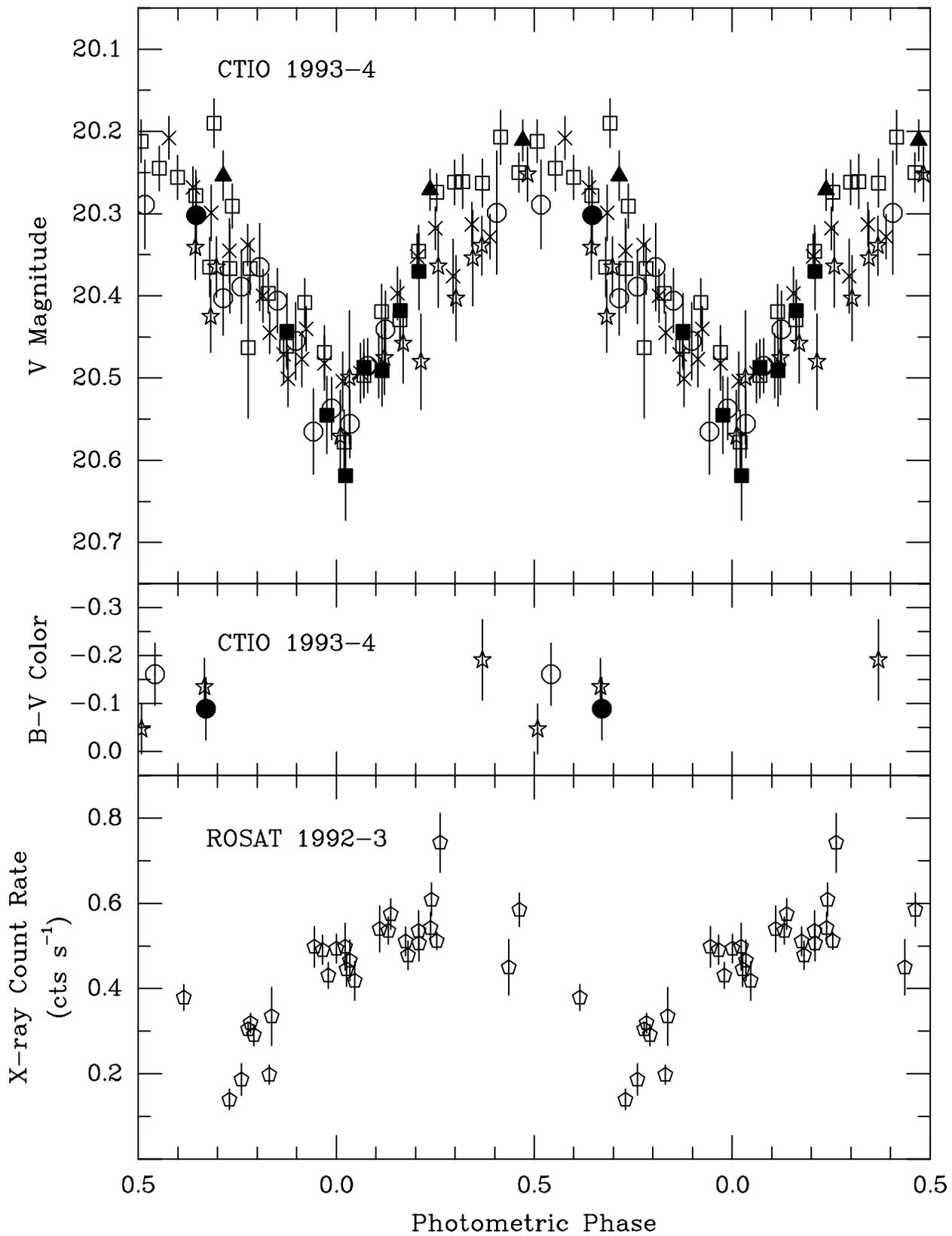

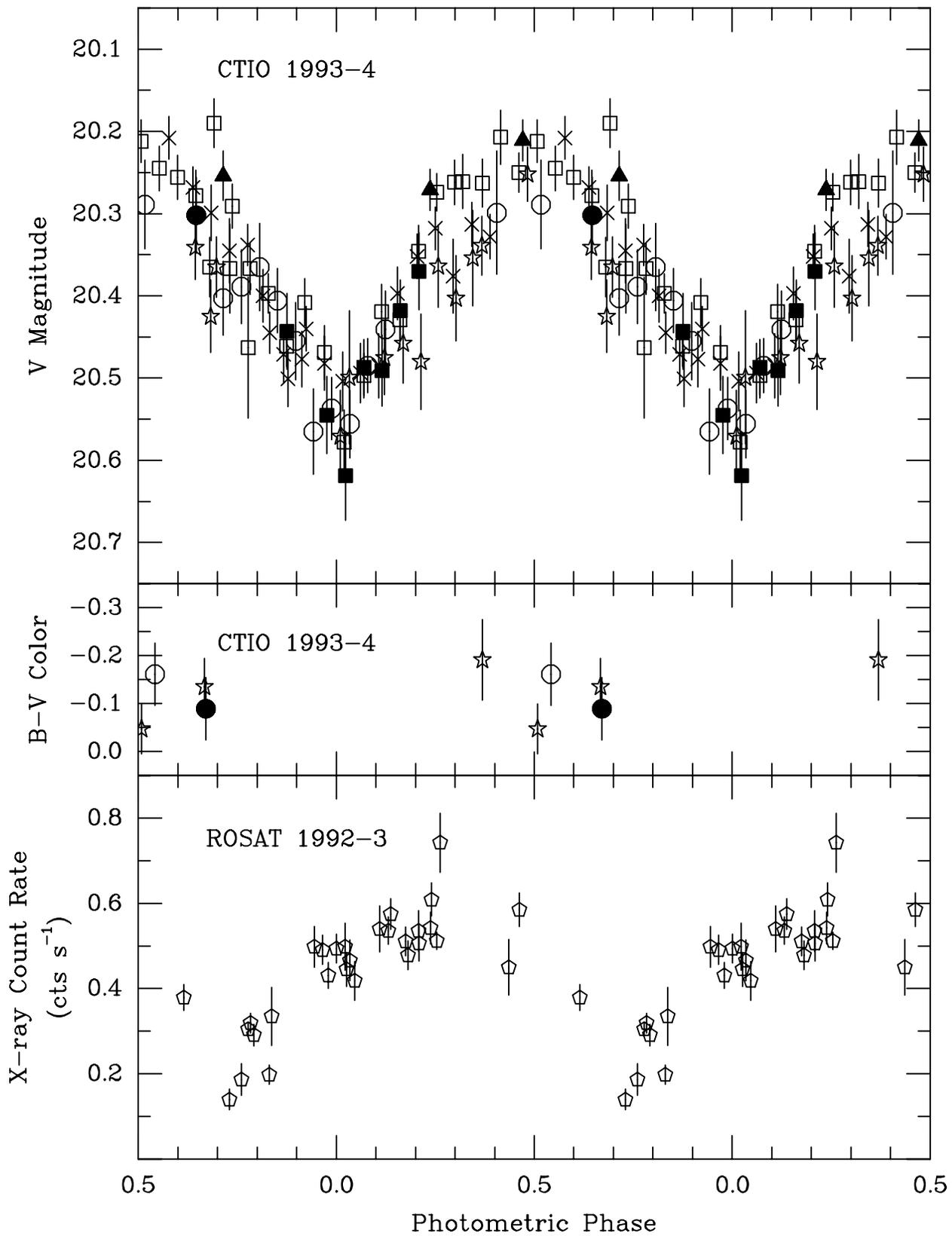

# A SUPERSOFT X-RAY BINARY IN THE SMALL MAGELLANIC CLOUD

P.C. Schmidtke[1], A.P. Cowley[1], and T.K. McGrath[1]

Department of Physics & Astronomy, Arizona State University, Tempe, AZ, 85287

e-mail: schmidtke@scorpius.la.asu.edu; cowley@vela.la.asu.edu; mcgrath@scorpius.la.asu.edu

J.B. Hutchings[1] and David Crampton[1]

Dominion Astrophysical Observatory, National Research Council of Canada, Victoria, B.C., V8X 4M6, Canada

e-mail: hutchings@dao.nrc.ca; crampton@dao.nrc.ca

## ABSTRACT

Photometric observations of the supersoft X-ray source 1E 0035.4−7230 obtained during two years reveal that the very blue optical counterpart ($V_{max} = 20.2$, $B - V = -0.15$, $U - B = -1.06$) undergoes nearly sinusoidal variations with a period of 0.1719256 days and an amplitude of $\Delta V \sim 0.3$ mag. *ROSAT* observations show the X-rays vary with approximately the same period. However, either the X-ray minimum precedes the optical minimum by about a quarter cycle or there is a small period difference between the two wavelength regions. We consider that this X-ray source is a close binary, with the optical light coming primarily from an accretion disk surrounding the compact star. Optical spectra show weak, variable He II (4686Å) emission which probably originates in this disk. Possible interpretations of the light curve are discussed, including X-ray heating of the secondary star. The very broad minimum in the X-ray light curve suggests the X-rays may be scattered in a large accretion disk corona (ADC) which is partially occulted, probably by an azimuthally irregular bulge on the disk rim. If this system lies at the distance of the Small Magellanic Cloud, it radiates near the Eddington luminosity.

*Subject headings:* accretion disks – close binaries; X-ray sources – stars: individual (1E 0035.4−7230)

## 1. INTRODUCTION

– 2 –

1E 0035.4−7230 was originally discovered as an X-ray source during an *Einstein* survey (Seward & Mitchell 1981) and later was included in a study of Small Magellanic Cloud sources by Wang & Wu (1992). This object is their source #13, so for simplicity we hereafter refer to 1E 0035.4−7230 as 'SMC 13'. Wang & Wu suggested that this source might be a low-mass X-ray binary (LMXB), noting its extremely soft X-ray spectrum which they compared to CAL 83 and CAL 87, both well-known "supersoft sources" (SSS) in the Large Magellanic Cloud (see review by Hasinger 1994). Jones et al. (1985) refer to a possible optical counterpart with $m_V \sim 21$, but since no coordinates or finding chart are given by these authors it is impossible to know if they refer to the same star discussed here.

The X-ray spectral properties of SMC 13 are described by Kahabka, Pietsch & Hasinger (1994). They find it to have a *ROSAT*-PSPC count rate of 0.38 ct s$^{-1}$ and a hardness ratio HR1/HR2$\sim -0.98$, indicating an extremely soft spectrum. They derive a blackbody temperature of kT$\sim$40 eV for an absorbing column of $N_H \sim 5 \times 10^{20}$ cm$^{-2}$. In addition they found the source varied by more than a factor of three on a time scale of a few days. If at the distance of the SMC, the data imply an X-ray luminosity of $\sim 10^{37}$ erg s$^{-1}$ in the band from 0.16 − 3.5 keV (Wang & Wu 1992; Kahabka, Pietsch, & Hasinger 1994).

The optical counterpart is a $20^{th}$ mag blue star, whose brightness varies periodically every $\sim$0.1719 days (Schmidtke et al. 1994b). The X-rays have been found to show approximately the same periodicity (Kahabka 1995). Below we describe the photometric and spectroscopic characteristics of SMC 13.

## 2. PHOTOMETRY

### 2.1. The X-ray Position and Optical Counterpart

Although an *Einstein* High Resolution Imager (HRI) position is available for SMC 13 (Wang & Wu 1992), improved accuracy can be obtained using *ROSAT* images. From our *ROSAT*-HRI data obtained on 1993 April 17 we derive an X-ray position, corrected for the 0.4° field rotation described by Kuerster (1993), of:

$$\text{X-ray position:} \quad \text{R.A.} = 00^h 37^m 20.3^s;$$
$$\text{Dec.} = -72°14'15'' \quad (\text{J2000.0})$$

Since this source is central in our image, its positional accuracy is $\sim \pm 5''$ (see Schmidtke et al. 1994a). The *ROSAT*-HRI count rate for SMC 13 during this observation was 0.085±0.008 ct s$^{-1}$.

The optical counterpart was independently identified with a faint blue variable star by Orio et al. (1994) and by Schmidtke et al. (1994a, see their Fig. 1). At maximum light the system has a magnitude of $V = 20.2$. A finding chart for SMC 13 is given in Fig. 1, with both the optical counterpart and the *ROSAT*-HRI position marked. Coordinates of the optical star, measured to an accuracy of $1''$, are:

$$\text{optical star:} \quad \text{R.A.} = 00^h37^m19.8^s;$$
$$\text{Dec.} = -72°14'13'' \quad (\text{J2000.0})$$

Thus, the optical counterpart lies $< 3''$ from the *ROSAT* X-ray position. The discovery of a common periodicity between the X-ray source and the optical star (see §2.3) definitely confirms the identification of the optical star with the X-ray source.

### 2.2. Photometric Observations

Photometry was obtained using the CTIO 0.9-m telescope with the TI#3 and Tek2048#2 CCDs during 1993 December 9–15 (UT) and 1994 November 3–10, respectively. A total of 91 direct $UBV$ images were taken, with integration times between 600 and 1200 s. The average seeing was $\sim 1''.4$, and the sky was photometric on most nights. These data were calibrated with observations of Landolt (1992) standard stars and reduced using DAOPHOT (Stetson 1987). For the $B$ and $V$ filters, differential magnitudes were calculated relative to local photometric standards within the CCD frames using our usual procedure which minimizes errors by PSF fitting (cf., Schmidtke 1988). However, the $U$ magnitudes are from the all-sky reductions since the local standard stars were red and did not have sufficient counts to calculate reliable differentials. All of the photometric data are presented in Table 1. The average errors are 0.061, 0.046, and 0.038 mag in $U$, $B$, and $V$, respectively. SMC 13 has a maximum magnitude near $V = 20.2$, with mean colors of $B - V = -0.15$ and $U - B = -1.06$.

### 2.3. Photometric Period

During the 1994 observing run it was obvious that the star was variable with a period near four hours. After reducing the data we formally searched for trial periods by calculating a periodogram (Horne and Baliunas 1986) using all $V$ magnitudes, including data from both 1993 and 1994 (see Fig. 2). Because the cycle count between these runs is uncertain, any one of 8–10 peaks in the periodogram could be the correct period. The periods with most power lie in the range from 0.1718 to 0.1722 days.

To choose among the likely periods, we have separately analyzed the *ROSAT* data taken in 1992 and 1993 during pointed observations of this source (Kahabka 1995). A periodogram of the X-ray count rate is shown by the lower curve in Fig. 2. There are two significant peaks in common with the optical periodogram, at frequencies corresponding to periods near 0.1719 and 0.1721 days. Using these values as starting points, we have refined the period using a least square fit to the optical data. We have also been guided by a plot of the *ROSAT* All Sky Survey (RASS) data which Kahabka kindly sent to us. Assuming that the minima in RASS data must be in phase with the *ROSAT* pointed data, we find that the best-fit period for SMC 13 is 0.1719256 days. Fig. 3 shows the optical and pointed X-ray data plotted on that period using the following photometric ephemeris:

$$T_0 = \text{HJD } 2449659.6001 \pm 0.0002$$
$$+ \ 0.1719256E \pm 0.0000005 \text{ days}$$

where $T_0$ is the mid-point of minimum $V$ light. The optical light curve appears to be rather smoothly modulated over a range of $\Delta V \sim 0.3$ mag with a broad, rounded maximum and a sharper minimum. The total width of the minimum is about half of the cycle. Some changes in the light curve from cycle to cycle may be present, but they are small compared to the overall brightness variation.

Also shown in Fig. 3 is the $B - V$ color curve plotted on the same period. Considering the errors in the colors, the curve is consistent with no variation in color with phase. However, we note that no color information was obtained near or at minimum light, when changes might be most pronounced. If one plots $B - V$ against $V$, there is a small indication that the system becomes bluer as the magnitude decreases, but this trend is based on only the five observations shown in Fig. 3 and may not be real. More two-color data with better phase coverage are needed to determine how the system's color changes with magnitude or phase.

The bottom panel in Fig. 3 shows the pointed X-ray data (Kahabka 1995) plotted with the same period and $T_0$ as the optical photometry. While the X-ray intensity curve shows the same general shape as the optical light curve (but with a larger amplitude), the time of X-ray minimum occurs about a quarter of a cycle earlier than the optical minimum. This is discussed further in §4. Kahabka (1995) has found that the X-ray count rate and hardness ratio increase together.

We note that the variations in $B$ light observed by Orio et al. (1994) in 1993 November (see their Fig. 2) are consistent with the period presented in this paper. However, their data were taken too close in time to our 1993 photometry to provide a strong discriminant between alias periods. From a single $V$ measurement, Orio et al. (1994) found $B - V = -0.29$ for this star, which is much bluer than any of our observations.

## 3. SPECTROSCOPY

Spectroscopy of SMC 13 was obtained with the CTIO 4-m telescope on three nights in 1994 November. The RC spectrograph with Reticon CCD was used with two different gratings giving ∼0.9Å per pixel covering the 3900–5020Å region and ∼1.8Å per pixel covering the 3720–5850Å region. Each observation of the star was preceded and followed by an observation of a He-Ar lamp while the telescope was at the target's position, so the wavelength scale is well established. One-dimensional spectra were extracted and processed following standard *IRAF* techniques to yield wavelength-calibrated spectra. Extremely variable seeing during the observing run made accurate flux calibrations impossible.

In 1994 a total of nine spectra were obtained. The two taken on the first night of the observing run (1994 November 10) with the higher dispersion grating are very underexposed; the remainder were taken with lower dispersion on 1994 November 12 and 13. The integration times of 45 min each were too long to give good phase resolution through the 4-hr period. Also, the observations are concentrated in only half of the cycle, between phases 0.45 to 0.98.

Five weakly exposed spectra were obtained 1993 December 12–14 using the CTIO multi-fiber system ARGUS on the 4-m telescope. These spectra covered the wavelength region from 3650–5800Å with a resolution of ∼1.8Å per pixel. We have co-added these 30-minute integrations to examine the average spectrum of SMC 13 during this time period.

The mean spectra of SMC 13 obtained from the seven low-resolution integrations taken in 1994 and the five spectra from 1993 are displayed in Fig. 4. SMC 13 shows a nearly continuous spectrum with only very marginal He II 4686Å emission present. An absorption feature near H$\beta$ is visible, but no other Balmer lines are detected. The 'emission feature' at ∼3955Å in the 1994 data appears to be an instrumental artifact, as it is not present on the two spectra taken with higher dispersion in 1994 nor in the 1993 spectra.

He II emission is easily detected on a few individual 1994 spectra, but is only marginally present on others (see Fig. 4). Thus, the line may be variable, but the low signal/noise of our spectra prevents us from measuring its strength reliably. The detectability of the line does not appear to be phase related. There is no evidence that any of the other emission lines reported by Orio et al. (1994) are present. They claimed O II (4077Å and 4118Å) and O I (4971Å) were stronger than He II, but neither our individual spectra nor the summed spectrum show such features. Furthermore, their other identifications seem doubtful based on the lack of additional lines which would be expected to be much stronger if these features were truly present. For example, N II 6065Å could not be present without much stronger lines from many other multiplets being visible. Even H$\beta$ emission is doubtful in their spectrum, based on the weakness of any emission near H$\alpha$. Thus, our spectra are consistent with the lower resolution one of Orio et al., with only He II 4686Å emission clearly present.

We have attemped to determine the He II velocity in the 1994 spectra in a variety of ways. Because of its weakness and the low signal/noise, the line is extremely difficult to measure. For those spectra in which the line is apparent, we measured its position by fitting a parabola. In addition, we cross-correlated the 1994 spectra in the region between $4500 - 5000$Å in each spectrum against the sum of all spectra for that observing run. The results are inconclusive. The velocity appears to change, but there is large scatter and no clear trend with phase for the small number of spectra available. The mean velocity of He II, obtained from the summed 1994 spectrum, is +33 km s$^{-1}$, considerably lower than expected for a SMC object (mean SMC velocity is +135 km s$^{-1}$ from planetary nebulae, according to Dopita et al. 1985). However, our spectroscopic data are biased in phase, as mentioned above. With such a short orbital period, one would expect to observe a large range in velocities, so that the low velocity measured from a few spectra does not rule out SMC membership. The sum of the 1993 spectra show an even weaker and broader feature at the He II line, barely distinguishable from the noise. By contrast, its velocity is $\sim +380$ km s$^{-1}$, much higher than the SMC velocity. Clearly, improved spectral data are needed to measure the systemic velocity and other orbital elements for SMC 13.

## 4. DISCUSSION

### 4.1. The Orbital Period of SMC 13

Although SMC 13 clearly undergoes periodic variations, their cause is not entirely obvious. We have first considered whether the true orbital period in SMC 13 is 0.1719 days or if it could be double that value. In some X-ray binaries with low-amplitude light curves, the orbital period is twice that found in the period analysis. Such double-peaked light curves generally are caused by ellipsoidal variations from the mass-losing star (e.g. A0620$-$00, McClintock & Remillard 1986, and Cen X-4, McClintock & Remillard 1990). In such cases the minima, and sometimes the maxima, have different depths, and generally the mass-losing star is seen in the spectrum. For SMC 13 there is no clear spectral evidence of the secondary star, but our spectra are of low quality. If one plots the $V$ light curve on twice the 0.1719-day period, the maxima and minima appear to differ somewhat, but the differences are within the errors and depend on results from nights of poorer seeing.

To resolve this possible ambiguity in the orbital period, we examined the *ROSAT*-PSPC data. When the X-ray count rates are plotted versus phase using double the 0.1719-day period, the curve becomes very scattered, varying by over a factor of three but with no simple trend in phase. The clear modulation of the X-ray data with P = 0.1719 days suggests this shorter period is the true orbital period of SMC 13.

### 4.2. Possible Interpretations of the Optical and X-ray Light Curves

The extreme width of the minima suggests that a simple eclipse of one star by the other cannot explain either the optical or X-ray light curves. The rounded maximum and sharper, extremely broad minimum of optical light curve are reminiscent of systems in which the inward facing hemisphere of the secondary star undergoes X-ray heating (e.g. Her X-1, Gerend & Boynton 1976, and 4U 2129+47, Thorstensen et al. 1979). In SMC 13 the observed amplitude of the optical variations is much smaller than in these other well-known X-ray heated systems. Furthermore, in such systems the spectrum of the heated star is visible, but in SMC 13 no stellar spectrum is apparent. However, if the accretion disk were the dominant source of light, then the light variations from the heated star would contribute only a small amount to the total light of the system, thus reducing the observed light amplitude and making the stellar spectrum less easily detected. Both the color and spectrum of SMC 13 indicate that we are primarily observing an accretion disk.

In a short period system the mass-losing star must be a G or K dwarf. Since the inner hemisphere of the star would be strongly heated, hydrogen Balmer lines would be expected. We carried out a simple test to see whether X-ray heating on such a secondary star would be obvious in a spectrum dominated by a much brighter accretion disk. We added the spectrum of an A star to that of SMC 13. Spectral features of the A star are no longer detectable when the A star contributes less than a quarter of the total light. This is consistent with the observed light amplitude of only $\sim$0.3 mag in a system where heating could cause a brightness difference between the two hemispheres of several magnitudes.

The difference in the phasing of the X-ray and optical minima could be caused by several different factors. There might be a small difference between the X-ray and optical periods due to a precessing disk or the presence of the third body, as has been found in a few systems (e.g. 4U 1915−05, Grindlay et al. 1988; Schmidtke 1988). Thus, comparing X-ray data obtained in 1992−93 with optical data from 1993−94 could result in an apparent difference in the phases of minima if periods in the two wavelength regions are not the same. Only when we are able to observe both regions simultaneously will we know whether the minima occur at the same time. If the periods are different, then the relative phasing of the X-ray and optical minima should change. However, if the minima are displaced by a constant phase shift, as shown in Fig. 3, then a single period is indicated.

We prefer an alternate interpretation for the phasing of the X-ray minima. Their long duration (nearly half the orbital cycle) indicates either the source being occulted or the object causing the eclipse (perhaps both) is spatially extended. The X-rays are probably scattered through a large accretion disk corona (ADC) similar to that found in X1822−37 (Mason & Cordova 1982; Hellier & Mason 1989). The X-ray mimima could be caused by partial obscuration of the ADC by a bulge in the disk where the mass transfer stream impacts it. Such an azimuthally irregular disk has been inferred in several systems and appears to account qualitatively for many aspects of their X-ray light curves (e.g. X1822−37, Hellier & Mason 1989).

Kahabka found that the hardness ratio of SMC 13 varies with the source brightness, with the spectrum being softest when the source is weakest. This indicates there is temperature structure in the disk, with the hardest (hottest) X-ray component located nearest to the compact star. This region would be hidden from the observer's view when the disk bulge is in conjunction with it.

Thus, in a qualitative sense, the observed optical and X-ray variations can be understood in terms of a model involving a X-ray heated secondary star and an irregular disk in which X-rays are both scattered and self-occulted. A number of tests of this picture should be made.

Simultaneous X-ray and optical photometry will show whether there is a phase offset between the two minima. Observing how the color varies with phase will allow us to verify whether heating of the secondary star is important. Higher signal/noise spectra may reveal the presence of the mass-losing star, especially at light maximum. Measuring velocity variations of the He II line will give us information about the motions in the disk and the orbit of the imbedded compact star. Long term variations in the light curves may reveal changes in the disk structure or disk precession. Better time resolution in the X-ray data will show if the inclination is high enough for the secondary star to briefly eclipse the central X-ray source, as in X1822−37, thus allowing us to place strong contraints on the orientation of the stars and the orbital inclination. Determination of the nature of the component stars must await these new observations.

### 4.3. Distance and Luminosity of SMC 13

It is of interest to consider whether SMC 13 is a member of the Small Magellanic Cloud. Adopting a distance modulus of 18.8 for the SMC (van den Bergh 1992), one obtains an absolute magnitude at maximum light of $M_V \sim +1.4$. This is close to the average absolute magnitude for low-mass X-ray binaries of $M_V = +1.2 \pm 1.1$ (van Paradijs 1983) and very comparable with systems of similar orbital period (van Paradijs & McClintock 1994). Thus, the optical luminosity of SMC 13 is entirely consistent with SMC membership. At this distance, its X-ray luminosity in the *Einstein* band is $L_x \sim 10^{37}$ erg s$^{-1}$ (Hasinger 1994), similar to three LMC supersoft X-ray binaries (CAL 83, CAL 87, and RX J0513.9−6951) all of which radiate at or near the Eddington luminosity.

We thank P. Kahabka for sharing his X-ray data with us. This work has been supported by both NSF and NASA. TKM's observing expenses were partially paid for by a gift from Sigma Xi.

– 9 –

– 10 –

Fig. 1.— Finding chart for 1E 0035.4−7230 with optical counterpart indicated and *ROSAT*-HRI position marked by a '+'.

Fig. 2.— Periodograms of data for 1E 0035.4−7230. The optical power spectrum is calculated using the $V$ magnitudes from Table 1, while the X-ray plot uses *ROSAT* data from Kahabka (1995).

Fig. 3.— Optical photometry and *ROSAT*-PSPC data of 1E 0035.4−7230 folded on the best-fit ephemeris (see text). (upper) The $V$ light curve with individual nights or runs shown by different symbols. (middle) The $B-V$ color curve. (lower) *ROSAT* data from pointed observations (Kahabka 1995). Note that the X-ray minimum occurs about a quarter of a cycle earlier than the optical minimum.

Fig. 4.— Spectra of 1E 0035.4−7230. (upper) Sum of two 1994 spectra in which He II emission was clearly visible. (middle) Mean of the seven low-dispersion spectra obtained in 1994 November. (lower) Mean of the three 'best' 1993 December spectra. The position of He II 4686Å is marked, although it is not distinguished from the noise in many individual spectra. The absorption at ∼4868 (H$\beta$?) is present on many spectra. We also mark the position of the O VI line (5290Å) which is sometimes found in spectra very hot stars, but does not appear to be present here.

TABLE 1
Photometry of 1E 0035.4−7230

| HJD (2440000+) | V | HJD (2440000+) | V | HJD (2440000+) | V |
|---|---|---|---|---|---|
| 9330.6176 | 20.252 | 9659.6509 | 20.376 | 9660.6831 | 20.262 |
| 9332.5430 | 20.425 | 9659.6589 | 20.313 | 9660.7486 | 20.365 |
| 9333.5680 | 20.341 | 9659.6668 | 20.328 | 9660.7573 | 20.367 |
| 9334.5520 | 20.339 | 9659.7402 | 20.400 | 9660.7652 | 20.463 |
| 9334.6088 | 20.365 | 9659.7492 | 20.471 | 9662.5635 | 20.271 |
| 9334.6624 | 20.571 | 9659.7571 | 20.477 | 9662.6038 | 20.211 |
| 9335.5413 | 20.475 | 9660.5145 | 20.261 | 9662.6455 | 20.254 |
| 9335.5494 | 20.458 | 9660.5232 | 20.263 | 9664.5646 | 20.443 |
| 9335.5571 | 20.480 | 9660.5311 | 20.207 | 9664.5818 | 20.545 |
| 9335.5645 | 20.364 | 9660.5390 | 20.250 | 9664.5902 | 20.619 |
| 9335.5723 | 20.403 | 9660.5469 | 20.212 | 9664.5980 | 20.487 |
| 9335.5797 | 20.354 | 9660.5548 | 20.245 | 9664.6060 | 20.491 |
| 9336.5576 | 20.499 | 9660.5627 | 20.256 | 9664.6139 | 20.418 |
| 9659.5274 | 20.208 | 9660.5706 | 20.278 | 9664.6220 | 20.370 |
| 9659.5378 | 20.268 | 9660.5786 | 20.190 | 9665.5152 | 20.299 |
| 9659.5458 | 20.299 | 9660.5865 | 20.291 | 9665.5345 | 20.289 |
| 9659.5537 | 20.345 | 9660.5944 | 20.367 | 9665.5683 | 20.403 |
| 9659.5616 | 20.338 | 9660.6023 | 20.397 | 9665.5762 | 20.389 |
| 9659.5713 | 20.445 | 9660.6102 | 20.461 | 9665.5843 | 20.365 |
| 9659.5792 | 20.501 | 9660.6180 | 20.408 | 9665.5920 | 20.406 |
| 9659.5870 | 20.440 | 9660.6265 | 20.469 | 9665.5999 | 20.455 |
| 9659.5949 | 20.482 | 9660.6352 | 20.578 | 9665.6077 | 20.565 |
| 9659.6028 | 20.504 | 9660.6436 | 20.497 | 9665.6156 | 20.537 |
| 9659.6106 | 20.494 | 9660.6515 | 20.419 | 9665.6234 | 20.556 |
| 9659.6185 | 20.485 | 9660.6594 | 20.429 | 9665.6311 | 20.485 |
| 9659.6268 | 20.397 | 9660.6674 | 20.346 | 9665.6387 | 20.441 |
| 9659.6352 | 20.352 | 9660.6753 | 20.274 | 9666.5884 | 20.302 |
| 9659.6430 | 20.318 | | | | |